\documentclass[
superscriptaddress,
 amsmath,amssymb,
 aps,
prl,
twocolumn,
]{revtex4-2}

\def\Thppp{\ensuremath{\textrm{Th}^{3+}}}

\newcommand{\thor}{\ensuremath{^{229}\textrm{Th}}}
\newcommand{\thorSP}{\ensuremath{^{229}\textrm{Th }}}

\newcommand{\lisaf}{\ensuremath{\textrm{LiSrAlF}_6}}

\newcommand{\caf}{\ensuremath{\textrm{CaF}_2}}
\def\thf{\ensuremath{\thor \textrm{F}_4}}

\newcommand{\thcaf}{\ensuremath{^{229}\textrm{Th:CaF}_2}}
\newcommand{\thlisaf}{\ensuremath{^{229}\textrm{Th:LiSrAlF}_6}}

\usepackage{graphicx}
\usepackage[colorlinks=true]{hyperref}
\usepackage{gensymb}
\usepackage{amsmath}
\usepackage{upgreek}
\usepackage{float}  
\usepackage[version=3]{mhchem}

\usepackage[dvipsnames]{xcolor}

\begin{document}

\title{Host-dependent frequency offsets in \ce{^229Th} nuclear clockwork}
\author{U. C. Perera}
\affiliation{Department of Physics, University of Nevada, Reno, Nevada 89557, USA}

\author{H. W. T. Morgan}
\affiliation{Department of Chemistry and Biochemistry, University of California, Los Angeles, Los Angeles, CA 90095, USA}
\affiliation{Department of Chemistry, University of Manchester, Oxford Road, Manchester M13 9PL, UK}

\author{Eric R. Hudson}
 \affiliation{Department of Physics and Astronomy, University of California, Los Angeles, CA 90095, USA}
 \affiliation{Challenge Institute for Quantum Computation, University of California Los Angeles, Los Angeles, CA, USA}
 \affiliation{Center for Quantum Science and Engineering, University of California Los Angeles, Los Angeles, CA, USA}
\author{Andrei Derevianko }
\affiliation{Department of Physics, University of Nevada, Reno, Nevada 89557, USA}
\date{\today} 

\begin{abstract}
Recent  advances in laser excitation of the low-energy nuclear isomer transition in \thor{}  have opened avenues for developing nuclear clocks, a novel quantum technology with exceptional performance and sensitivity to exotic physics. Here we explore the host-dependence of the nuclear clock frequency, focusing on the isomer shift induced by the difference in the nuclear charge distribution between the ground and excited nuclear states. We combine relativistic many-body  methods of atomic structure with periodic density functional theory to evaluate the isomer shifts in solid-state hosts. We elucidate the critical importance of the ``relaxation'' effect in evaluating the isomer shifts.
Our analysis predicts nuclear clock frequencies for various solid-state and trapped ion platforms: $ \omega_\text{clk}(\text{solid state}) =  2,020,407,384(40) \, \text{MHz}$, $\omega_\text{clk}(\thor^{4+})  =  2,020,407,648(70) \, \text{MHz}$, and $ \omega_\text{clk}(\thor^{3+}) =  2,020,407,114(70) \, \text{MHz}$. 
We also determine the nuclear transition energy for the bare \thor{} nucleus to be $\omega_\text{nuc} = 8.272(22) \,\text{eV}$. 
Our calculated valence-band isomer shifts for different host materials constrain the nuclear transition frequencies to an 80 MHz-wide frequency window, aiding experimental searches for the \thor{} nuclear transition in novel materials. 
\end{abstract}

\maketitle
Recent observations~\cite{Tiedau2024,Elwell2024,Zhang2024-Th229Comb,zhang2024thf} of laser excitation of a uniquely low-energy nuclear isomer transition in \thor{} pave the road for realizing a novel class of timekeeping devices, nuclear clocks. Beyond timekeeping, this nascent quantum technology can be exquisitely sensitive to a variety of exotic physics~\cite{SafBudDeM2018.RMP}.  One may distinguish between two nuclear clock platforms~\cite{PieTam03}\footnote{Clocks based on neutral Th atoms would suffer from the rapid internal conversion process whereby the \thorSP nucleus transfers its excitation to the atomic electrons, ionizing the Th atom.}: (i) trapped Th ions, e.g., \Thppp{}~\cite{CamRadKuz12}  and (ii) \thor-containing crystals~\cite{Rellergert2010}, such as \thor:\lisaf, \thor:\caf{} doped bulk crystals~\cite{Tiedau2024,Elwell2024,Zhang2024-Th229Comb}, or \thf{} films~\cite{zhang2024thf}. The two platforms offer distinct trade-offs between the nuclear clock accuracy and stability. 
{
Solid‐state  clocks offer superior stability by interrogating macroscopic ensembles of nuclei simultaneously and provide a pathway to truly portable, high‐performance timekeeping. 
However, solid‐state platforms are prone to inhomogeneous broadening and environment‐dependent shifts, though recent work may show ways to largely mitigate these~\cite{Morgan2025-design-of-Th-materials,morgan2025-spinless}.
By contrast, the trapped‐ion platforms, which should provide superior accuracy, require long integration times to reach their ultimate resolution.

It can thus be expected that both types of nuclear clocks will find great use and it is important to understand their relationship. 
In particular, the nuclear transition energy, which defines the ``ticking'' rate of the clock, can be expected to vary due to the interaction of nuclei with the electronic degrees of freedom in various hosts. Here we evaluate isomer shifts induced by the difference in the nuclear charge distribution between the ground and excited nuclear states. These primarily govern differences (offsets) in nuclear clock frequencies between various \thor{} hosts.
Our analysis predicts nuclear clock frequencies for these platforms, which both aids the experiments and relates the results for the two classes of nuclear clocks allowing them to be refined together.}

The nuclear clock transition frequency $\omega_\mathrm{clk}$ is affected by the host environment of the \thor{} nucleus. So far the nuclear transition signal has been observed only in the solid-state platforms with the measured transition frequencies~\cite{zhang2024thf,Zhang2024-Th229Comb,Elwell2024}:
$\omega_\mathrm{clk}(\thf) = 2,020,407.4(3)_\mathrm{stat} (30)_\mathrm{sys} \, \mathrm{GHz}$, $\omega_\mathrm{clk}(\thlisaf) =2,020,407.3(5)_{\textrm{stat}}(30)_{\textrm{sys}} \, \mathrm{GHz}$, and $\omega_\text{clk}(\thcaf) = 2,020,407,384.335(2) \, \text{MHz}$. 
While the already measured frequencies agree at the stated 30~GHz level of accuracy, we show that a difference arises below roughly 100~MHz resolution, well above the demonstrated kHz VUV frequency comb resolution~\cite{Zhang2024-Th229Comb}.

There is a wide variety of other hosting materials proposed~\cite{Morgan2025-Th229NonLinear,morgan2025-spinless,Morgan2025-design-of-Th-materials,Ellis2014,RN552}. A practical challenge faced by an experimentalist is to pinpoint the narrow nuclear transition by scanning the laser frequency.  We simplify this tedious process by exacting a $\sim 100 \,\mathrm{MHz}$-wide scanning window for the nuclear clock spectroscopy. We predict the nuclear clock frequencies for both the solid state and trapped ion platforms:
\begin{align}
 \omega_\text{clk}(\text{solid state}) = & 2,020,407,384(40) \, \text{MHz} \,, \label{Eq:Solid}\\
 \omega_\text{clk}(\thor^{4+})  = & 2,020,407,648(70) \, \text{MHz} \,, \label{Eq:Th4plus} \\
 \omega_\text{clk}(\thor^{3+},5f_{5/2}) = & 2,020,407,114(70) \, \text{MHz} \,. \label{Eq:Th3plus}
\end{align}
We also deduce the nuclear transition energy in the bare \thor{} nucleus (i.e., the \thor$^{90+}$ ion)
\begin{equation}
  \omega_\text{nuc} = 8.272(22) \,\text{eV} \,.    \label{Eq:ThBare}
\end{equation}

To this end, we examine how the nuclear clock frequency depends on the local electronic environment of the \thorSP nucleus. This dependence arises from the isomer (chemical) shift, which results from the differences in the nuclear charge distributions between the nuclear ground and isomeric states. Remarkably~\cite{morgan2025-spinless},  a competing second-order Doppler shift is significantly reduced for \thor{} compared to nuclear transitions used in M{\"o}{\ss}bauer spectroscopy because the nuclear transition energy in \thor{}, Eq.~\eqref{Eq:ThBare}, is several orders of magnitude smaller. Otherwise, the chemical shifts are often masked by the Doppler shifts~\cite{gibb1976-MossbauerBook}. We develop a method to evaluate the isomer shift in solid-state hosts. While compatible with periodic density functional theory (DFT) approaches, our method incorporates essential relaxation effects that are difficult to account for in DFT in a relativistic framework required for heavy thorium.

The nuclear transition of interest is nominally between the ground $|g\rangle=|(5/2)^+[633]\rangle$ and the isomer $|e\rangle =|(3/2)^+[631]\rangle$ nuclear states of \thor, where we use the Nilsson quantum numbers. Since the two nuclear states possess different charge distributions and electromagnetic moments interacting with electrons, the clock frequency is determined by the eigenenergies  of the {\em compound} nucleus-electron system. Namely, transitions between the eigenstates of the compound system are probed in laser spectroscopy. We  distinguish between the monopole (electric charge distribution), magnetic dipole, and electric quadrupole interactions of the nucleus with the electrons. The effects of the magnetic and quadrupolar couplings interacting with the local electronic environment can be removed by proper weighted averaging of transition frequencies between various magnetic sublevels~\cite{Zhang2024-Th229Comb}. It is the  resulting  frequency that we refer to as the clock frequency $\omega_\mathrm{clk}$. This averaging cannot, however, remove the effects of the electrons interacting with the nuclear charge distributions due to the scalar nature of the interaction and $\omega_\mathrm{clk}$ depends on the electronic environment (leading to the isomer shift $ \delta \omega_\mathrm{iso}$ of the clock frequency). 

{\em Formalism ---} We start with the conventional parameterization~\cite{Filatov2009-review} of the isomer shift (also known as the chemical shift) in M{\"o}{\ss}bauer spectroscopy and later demonstrate its inadequacy for the \thor{} solid-state hosts. The conventional parameterization reads
 \begin{equation}
  \delta E_\mathrm{iso} = \frac{4 \pi}{5} Z S(Z)  R^2 \frac{\Delta R}{R} \rho(0) \,, \label{Eq:SimpleIsomerShift}
 \end{equation}
where $Z$ and $R$ are the nuclear charge and  the r.m.s charge radius, respectively, $\Delta R = R_e - R_g$ is the difference in the r.m.s nuclear charge radii for the isomer and the ground states, and  $\rho(0)$ is the electron density at the nucleus. In practice, considering the complexities of the solid state environment, $\rho(0)$  is replaced with the non-relativistically calculated electronic density at the nucleus. The factor $S(Z)$ accounts for the effects of relativity and  non-uniformity of electron density inside the nucleus. Eq.~\eqref{Eq:SimpleIsomerShift} is derived by computing the expectation value of 
$\delta V(\mathbf{r}) = \frac{3}{2} \frac{Z}{R} \frac{\Delta R}{R} \left( 1 - (r/R)^2 \right)$ inside the nucleus. This derivation, among other factors~\cite{Filatov2009-review},  misses a critically important relaxation effect, as discussed below.

For \thor,  $R_{g} = 5.756(14) \, \mathrm{fm}$  \cite{ANGELI201369} and the  difference in the nuclear charge radii~\cite{Yamaguchi2024} $\delta \langle r^2 \rangle^{229m,229} = 0.0097(26) \, \mathrm{fm}^2$, leading to  $\Delta R =  0.00084(23)  \, \mathrm{fm}$.
Indirect determinations~\cite{SafPorKoz2018-ThIsomerShift,DzubaFlam2023-ThIsomerShift} of $\delta \langle r^2 \rangle^{229m,229}$ are consistent with Ref.~\cite{Yamaguchi2024}; we will use the results of the {\em direct} determination~\cite{Yamaguchi2024} as a reference value throughout. One can convert our results to the field shifts $F \equiv \delta E_\mathrm{iso}/\delta \langle r^2 \rangle^{229,229m}$ used in the isotope shift literature. 

The isomer shifts in  \thor{} ions  can be reliably computed using relativistic   many-body atomic structure methods~\cite{SafPorKoz2018-ThIsomerShift,DzubaFlam2023-ThIsomerShift}.   \thor{} isomer shifts in solid-state hosts, the focus of our paper, have not been evaluated so far.  

We start with a discussion of validity of Eq.~\eqref{Eq:SimpleIsomerShift} by testing the underlying assumptions for the Th$^{3+}$ ion.  Our computed first-order in $\delta V$  correction  to the  energy of the Th$^{3+}$ ground state valence electron is $\delta \varepsilon_{5f_{5/2}}^\mathrm{(1),iso} = 4  \times 10^{-12} \, \mathrm{MHz}$.  However, running  the Dirac-Hartree-Fock (DHF) code~\cite{WRJBook} for Th$^{3+}$ with different $R$ and taking the difference of the DHF energies  yields  $\delta \varepsilon_{5f_{5/2}}^\mathrm{iso} = -5 \times 10^2 \, \mathrm{MHz}$, a 14 orders (!) of magnitude difference (see the Supplemental Materials (SM) for further details).

The reasons behind the gross inadequacy of the expression for $\delta V$ have been elucidated in Ref.~\cite{DerRavJoh04} for a similar problem of determining QED vacuum polarization corrections, where the nuclear potential is modified by the short-ranged Uehling potential. 
Briefly, the change in the nuclear potential mostly affects the deeply bound core electrons that have a large overlap with the nucleus. This, in turn, affects the self-consistent DHF electrostatic potential $U(\mathbf{r}|R)$ due to the core electrons, which has a much larger effect on the energies of $\ell \neq 0$ states~\footnote{Relativistically, $j>1/2$ orbitals} than $\delta V$ due to the bare nucleus. Thereby, following~\cite{DerRavJoh04}, the ``dressed'' nuclear perturbation  is rather,    $ \delta\tilde{V} = \delta V  + U(\mathbf{r}|R + \Delta R) - U(\mathbf{r}|R)$.  Then the range of  $ \delta V$ is effectively increased from $R$ to $\sim a_0/Z$, where $a_0$ is the Bohr radius, see Fig.~2 in Ref.~\cite{DerRavJoh04}. $a_0/Z$ is the spatial extent of the $1s_{1/2}$  and $2p_{1/2}$  core shells. Moreover, we can simply compute the isomer shift as the expectation value of this ``dressed'' perturbation, so that the isomer shift for the $n \ell m$ state of the 
\thor$^{3+}$  ion is (see also Ref.~\cite{DzubaFlam2023-ThIsomerShift})
\begin{equation}
\delta \varepsilon_{n\ell}^\mathrm{iso}(\thor^{3+}) = \langle n \ell m | \delta \tilde{V} |   n \ell m \rangle  \,. \label{Eq:IsoShiftExp}
\end{equation}
 {
As a side note, the label ``relaxation'', while commonly used in the atomic physics community may be somewhat misleading to the general reader as it suggests a time-dependent process. Although a nuclear transition does indeed induce a readjustment of the electronic cloud, the timescale of this response — on the order of a fraction of an attosecond — is vastly shorter than the duration of any realistic clock-interrogation laser pulse. This justifies treating the electronic degrees of freedom as ``instantaneously'' following the change in nuclear charge distribution.
}

 In the valence band (VB) of the Th-doped high-bandgap ionic insulators and \thf,  Th  occurs in the +4 charge state~\cite{Dessovic2014,zhang2024thf,Morgan2025_internal_conversion,Elwell2024}. This suggests a qualitative picture~\cite{Morgan2025_internal_conversion} of the VB electrons scattering off the closed-shell Th$^{4+}$ ion and we may separate out the Th$^{4+}$ ion contribution common to all the hosts: 
\begin{align}
  \delta E_\mathrm{iso} =  \delta E_\mathrm{iso}(\text{Th}^{4+}) + \delta E_\mathrm{iso} (\text{VB}).   
\end{align}
This separation is  supported by the electrons in closed-shell Th$^{4+}$ ion being deeply bound, leading to the suppression of core polarization effects. This is identical to the commonly-used frozen core approximation in atomic physics~\cite{WRJBook}. Thereby, the relative isomer shifts between the hosting materials is given by the differences in VB contributions, $\delta \omega_\mathrm{iso} (\text{VB})$. Notice that even if we were to solve the many-body problem exactly, the $\sim 30\%$ error in $\Delta R$ still would imply the same $30\%$ error in the isomer shift. Therefore, in the spirit of Ref.~\cite{Morgan2025_internal_conversion}, we develop a model that takes into account  major qualitative effects in these systems: we combine the widely used non-relativistic methods of periodic density-functional theory (DFT) to compute the VB properties and the relativistic atomic structure methods that  accounts for the important relaxation effect.

To determine the isomer shift, we use the mean-field approximation where the electrons can be treated as moving independently in a crystal potential. For concreteness,  in our numerical estimates, we use plane-wave DFT with periodic boundary conditions. 
For Th-doped crystals, such calculations are carried out with a periodically continued supercell containing several unit cells of the host material and a Th doping complex, which includes $\text{Th}^{4+}$ and  also other ions (such as F$^-$) or vacancies for charge compensation. For crystals such as \thf, where Th is a natural constituent, the calculations are carried out for a single unit cell.

{
Our derivation of the VB isomer shift contribution for the \thor{} solid-state clocks is presented in the SM. Briefly, we compute the expectation value of $\delta \tilde{V}$ in the valence band. Since the matrix elements  of $\delta \tilde{V}$ is accumulated near the Th nucleus ($r \lesssim a_0/Z$), we peel off the Th-centered contribution to the Bloch wavefunctions, by  expanding  it over the $\mathrm{Th}^{3+}$ ion atomic valence orbitals~\cite{Morgan2025_internal_conversion}. This brings in the Th-projected density of states (PDOS) $\rho_\ell(\varepsilon)$ for the lowest-energy Th$^{3+}$ valence orbitals of orbital angular momentum $\ell$. This provides a connection between atomic isomer shifts in \thor$^{3+}$, $\delta \varepsilon_\ell^\mathrm{iso}(\thor^{3+})$, and crystal properties:
\begin{equation}
\delta E_\mathrm{iso} (\text{VB})=
 \sum_\ell \text{IPDOS}_\ell  \times  \delta \varepsilon_\ell^\mathrm{iso}(\thor^{3+})  
 \,. \label{Eq:VB-iso-konec}
\end{equation}
Here $\text{IPDOS}_\ell  \equiv \int_\text{VB} \rho_\ell(\varepsilon) d \varepsilon$ is the integrated projected density of states (IPDOS) for orbital momentum $\ell$;  the  integration extends from the VB minimum to the VB maximum.    PDOSs   can be routinely computed in periodic DFT numerical packages, such as Vienna Ab initio Simulation Package~\cite{VASP} (VASP). As an example, detailed PDOS plots for Th:\lisaf{} can be found in Ref.~\cite{Morgan2025_internal_conversion}, see also Fig.~1 in the SM.
 
Eq.~\eqref{Eq:VB-iso-konec} can be  interpreted in the following way: the valence band isomer shift is a linear combination of isomer shifts of the \thor$^{3+}$ valence orbitals weighted by the VB occupation numbers of these orbitals. Indeed, the IPDOS yields the occupation number for a given $\mathrm{Th}$ valence orbital in a crystal.
In a chemical sense, Th is in the +4 oxidation state, so there are no valence electrons localized on Th - the non-zero IPDOS comes from overlap between thorium and neighbouring anions.
}

{\em Numerical results ---}
Computation of the IPDOS begins with the projected density of states (PDOS) produced by VASP, calculated with the MBJ functional, where the projection is done using the projector functions in the PAW scheme~\cite{Blochl1994}.
We then numerically integrate the Th PDOS, resolved by the $\ell$ orbital angular momentum ($s$-$p$-$d$-$f$).
The integration is done over the part of the valence band that describes anion-cation interactions.
The exact energy range varies between compounds but is typically from around -10 eV up to the Fermi level.
The bottom is always well defined by a large gap in the DOS.
The integration does not include the semi-core orbitals treated in the valence space by VASP (\textit{e.g.} Th 6$p$) because these electrons are covered in the DHF calculations.

Our computed IPDOS and valence band isomer shifts  are compiled in Table~\ref{Tab:IPDOS-and-iso} for the materials of current interest to the solid-state nuclear clocks. 

\begin{table}[h]
\centering
\caption{Valence band isomer shifts $\delta \omega_\mathrm{iso}(\text{VB})$ for various \thor{}-containing solid-state hosts and sites. The table also lists integrated partial densities of states $\text{IPDOS}_\ell$ for various partial waves.
}
\begin{tabular}{lcccccc}
\hline\hline
  $\text{IPDOS}_\ell$:  & $ s$ & $p$ & $d$ & $f$  & $\delta \omega_\mathrm{iso}(\text{VB})$, MHz \\
\hline
\multicolumn{6}{c}{\textbf{\thor:\lisaf}} \\
{GM 4x4x2} & 0.083 & 0.350 & 0.609 & 0.328    & -234\\
{GM 3x3x2} & 0.090 & 0.400 & 0.677 & 0.377      & -269\\
{LM1 3x3x2} & 0.092 & 0.431 & 0.683 & 0.405  & -284\\[1ex]
\multicolumn{6}{c}{\textbf{\thor:\caf}}\\
F-$90^\circ$-F  & 0.043 & 0.395 & 0.610 & 0.315 & -264 \\
F-$180^\circ$-F & 0.091 & 0.407 & 0.603 &  0.425 & -278 \\[1ex]
\multicolumn{6}{c}{\textbf{\thf}}\\
site 1 & 0.073 & 0.413 & 0.593 & 0.376 & -266\\
site 2 & 0.074 & 0.424 & 0.599 & 0.382 & -270 \\[1ex] 
\multicolumn{6}{c}{\textbf{\thor:BaMgF$_4$}}\\
2Fi  & 0.103 & 0.397 & 0.636 & 0.427 & -275\\
VMg  & 0.102 & 0.415 & 0.629 & 0.430 & -276\\
VBa  & 0.100 & 0.407 & 0.623 & 0.435 & -279\\[1ex] 
\multicolumn{6}{c}{\textbf{\thor:BaZnF$_4$}}\\
2Fi  & 0.101 & 0.413 & 0.637 & 0.452 & -290\\
VZn  & 0.088 & 0.337 & 0.562 & 0.373 & -243\\
VBa  & 0.090 & 0.375 & 0.581 & 0.389 & -254 \\
\hline\hline
\end{tabular}
\label{Tab:IPDOS-and-iso}
\end{table}

To compute the isomer shifts~\eqref{Eq:VB-iso-konec}, we combined the listed IPDOSs  with non-relativistic isomer shifts $\delta \varepsilon_\ell^\mathrm{iso}(\thor^{3+})$. To determine these, we took a weighted average of the isomer shifts of fine-structure components from Ref.~\cite{DzubaFlam2023-ThIsomerShift}, so that 
$  \varepsilon^\mathrm{iso}_{n\ell} = \left(  
  \ell\varepsilon^\mathrm{iso}_{n,\ell, j=\ell-1/2} +
 (\ell +1) \varepsilon^\mathrm{iso}_{n,\ell, j=\ell+1/2}\right)/(2\ell+1)
$. The weights are determined by the total number of magnetic substates for a given $j=\ell \pm 1/2$. 
The resulting $\delta \varepsilon_\ell^\mathrm{iso}(\thor^{3+})$  are  893, -25.5, -210, and -522 MHz  for $\ell=\overline{0,3}$. These shifts have a fully correlated 27\% error due to the uncertainty in $\Delta R$, implying  the minimum 27\% error in the isomer shifts. 

For \lisaf{} and \caf{} we included two lowest-energy \thor{} defects. For \lisaf, the GM (global minimum) defect formulation is \ce{Th^{..}_{Sr} + 2F_{i}$'$} and the LM1 (first local minimum) defect is \ce{Th^{..}_{Sr} + F_{i}$'$ + V_{Li}$'$}. Here we use the Kröger–Vink notation~\cite{kroger1974-book}.
For \caf{}, the lowest two defects both have the composition \ce{Th^{..}_{Ca} + 2F_{i}$'$} but differ in the orientations of the interstitial F$^-$. The lowest-energy structure has a F-Th-F angle of approximately 90\degree, while the next-lowest has an angle of 180\degree. These highly symmetrical arrangements are permitted by the cubic \ce{CaF2} structure. For \thf{}, there are two distinct Th sites in a unit cell, having different electric field gradients (EFG)~\cite{RN566}. The site of type 1 (4 atoms per unit cell) has a slightly larger EFG than the site of type 2 (2 atoms per unit cell) according to DFT~\cite{zhang2024thf}. We also include the isomer shift for several \thor{} doping sites in   frequency-doubling non-linear optical crystals,
BaMgF$_4$ and BaZnF$_4$. These crystals hold an intriguing promise for realizing a truly compact solid-state nuclear clock integrated with solid-state clock lasers~\cite{Morgan2025-Th229NonLinear}. Details of our calculations can be found in the SM.

Analysis of individual partial wave contributions to the isomer shift~\eqref{Eq:VB-iso-konec}
shows that the bulk of the sum comes from the $\ell \neq 0$ Th orbitals having a negligible overlap with the nucleus,  illustrating the critical importance of the relaxation effect. The conventional M{\"o}{\ss}bauer spectroscopy formula, Eq.~\eqref{Eq:SimpleIsomerShift}, is an {\em approximation}. As follows from the derivation, it works well only if the isomer shift is dominated by the $\ell = 0$ state contributions, as non-relativistically only these $s$ states have a non-vanishing electronic density $\rho(0)$ at the nucleus.
For example, for 
the \thor:\lisaf{} GM site computed with the $4\times 4\times 2$ supercell, the 
$\text{IPDOS}_\ell  \times  \delta \varepsilon_\ell^\mathrm{iso}(\thor^{3+})$  contributions are (in MHz) : $74.2$, $-8.93$, $-128$, and $-171$ for $\ell = \overline{0,3}$ with the $-234\, \text{MHz}$ total VB isomer shift. 
Neglecting the relaxation effect would yield substantially different results. Indeed, keeping only the $\ell=0$ contribution, as per Eq.~\eqref{Eq:SimpleIsomerShift}, for this GM site results in $\delta E_\mathrm{iso} (\text{VB}, \ell=0)= 74.2 \, \text{MHz}$, while the total VB isomer shift, including all the contributions, is $-234\, \text{MHz}$. We also point out that the Ref.~\cite{Higgins2025-Temperature-dependence-ThCaF2} interpretation of the temperature-dependent \thcaf{} isomer shift in terms of the change in $\rho(0)$ is inadequate because it is based on Eq.~\eqref{Eq:SimpleIsomerShift}.

{\em Implications --- }
Table~\ref{Tab:IPDOS-and-iso} indicates the valence band contribution of about -250 MHz ($\sim \text{meV}$) to the isomer shift in solid-state hosts. We find a $\sim 60$ MHz spread in $\delta E_\mathrm{iso}(\text{VB})$ over the considered hosts and doping sites. The 30\% uncertainty in $\Delta R$ can widen this spread to $\sim 80$ MHz. {Table~\ref{Tab:IPDOS-and-iso} covers a substantially broad class of hosting materials.} This implies that the high-resolution lasers need to be scanned over the $\sim 80$ MHz window when searching for the \thor{} transitions in novel solid state hosts. 
Taking the measured~\cite{Zhang2024-Th229Comb} $\omega_\text{clk}$ in \thcaf (F-$90^\circ$-F doping site) as the reference value, we arrive at  the estimated clock frequency in solid state hosts $\omega_\text{clk}(\text{solid state})$, Eq.~\eqref{Eq:Solid}.
 Notice that our $\omega_\text{clk}(\text{solid state})$ prediction is consistent with the non-observation of the isomer shift at the 30 GHz accuracy level in three hosts.  Yet the $80$ MHz window is several orders of magnitude larger than the currently demonstrated~\cite{Zhang2024-Th229Comb} $\sim\,\text{kHz}$ VUV frequency comb resolution, making the isomer shift not only discernible, but practically important for high-resolution nuclear spectroscopy. 

The isomer shifts of Table~\ref{Tab:IPDOS-and-iso}  can be used to determine the  frequency offsets between pairs of nuclear-clock materials, e.g., between \thlisaf{} and the already measured~\cite{Zhang2024-Th229Comb} $\omega_\text{clk}(\thcaf)$;
 this enables pinpointing laser frequency offsets with a $\sim 10 \, \text{MHz}$ accuracy. 

Of special interest are the nuclear clock frequencies in various \thor{} ion hosts, due to their projected $10^{-19}$ fractional inaccuracy~\cite{CamRadKuz12}. Removing $\delta E_\mathrm{iso}(\text{VB})=-264(70) \, \text{MHz}$ 
from $\omega_\text{clk}(\thcaf)$
yields the frequency in the \thor$^{4+}$ ion host, $\omega_\text{clk}(\thor^{4+})$ listed in Eq.~\eqref{Eq:Th4plus}.
From here, we predict the proposed \thor$^{3+}$ ground $5f_{5/2}$ state nuclear clockwork~\cite{CamRadKuz12} frequency, $\omega_\text{clk}(\thor^{3+},5f_{5/2})$, Eq.~\eqref{Eq:Th3plus}, 
by adding $\delta \varepsilon^\text{iso}_{5f_{5/2}} = -534 \, \text{MHz}$ to $\omega_\text{clk}(\thor^{4+})$.

{
Our calculation of isomer shifts in \thor{} ions is presented in the SM. Briefly, we run the atomic Dirac-Hartree-Fock code~\cite{WRJBook} with the Fermi nuclear distribution, vary its half‐density radius $c$, compute electronic energies $E(c)$, extract $dE(c)/dc$, and multiply by the $\Delta R$-implied 
difference $\Delta c$ between the half-density radii of the nuclear isomer and ground states. These results are consistent with earlier calculations~\cite{DzubaFlam2023-ThIsomerShift}.
}

The measured \thor{} transition frequencies in crystals can be related to that in the bare (\thor$^{90+}$) nucleus by subtracting off  $\delta E_\mathrm{iso}(\thor^{4+}) =  0.083(22) \, \text{eV}$, the isomer shift for the Rn-like \thor{} ion, see the SM. As we have demonstrated, the VB contributes at the meV level and can be neglected at this level of accuracy.  Thereby, we find that the measured $8.3557 \ldots\, \mathrm{eV}$ transition frequency in \thor-containing crystals~\cite{Tiedau2024,Elwell2024,Zhang2024-Th229Comb,zhang2024thf}  implies  $
\omega_\text{nuc} = 8.272(22) \,\text{eV} $
in the bare \thor{} nucleus, c.f. Eq.~\eqref{Eq:ThBare}.

The same crystal can host \thor{} at different lattice or doping sites. We find that in the \thf{} crystal~\cite{zhang2024thf}, the isomer shift between the two nonequivalent sites is 4(1) MHz, see Table~\ref{Tab:IPDOS-and-iso} and the SM. The \thor{}-doped high bandgap insulators exhibit much larger isomer shift variations across different sites. This can be used to spectroscopically separate the populations and also to test our estimates. For example, the VUV comb experiment~\cite{Zhang2024-Th229Comb} in \thcaf{} has focused on the nuclear transitions at the F-$90^\circ$-F doping site. This experiment has also detected some of the lines from the sub-dominant  F-$180^\circ$-F site. Further experiments would be beneficial to determine the central frequency for this sub-dominant site. Our prediction for the frequency offset between the two doping sites in \caf{} is 14(4) MHz.

Table~\ref{Tab:IPDOS-and-iso} lists our results  for the same \ce{Th^{..}_{Sr} + 2F_{i}$'$} GM doping site in \lisaf{} but for  two different supercell volumes. The  $3\times 3\times 2$ and $4\times 4\times 2$ supercells correspond to $n_\text{Th}$ Th doping concentrations of  $2.7 \times 10^{20}\, \mathrm{cm}^{-3}$ and $1.5 \times 10^{20}\, \mathrm{cm}^{-3}$, respectively.  Qualitatively, the isomer shift dependence on  $n_\text{Th}$ can be explained by an interaction of the defect with its 
neighboring defects, affecting the electron densities. In DFT, these arise as interactions of the defect with its periodically continued supercell images. We mention lattice deformations and the electric dipole-dipole interaction as potential sources of the isomer shift dependence on $n_\text{Th}$. The latter is due to the fact that a doping complex consists of Th$^{4+}$ surrounded by two F$^-$ interstitials. Such complex in a non-linear arrangement has a non-vanishing electric-dipole moment. While increasing the supercell size beyond $4\times 4\times 2$ in our DFT calculations is computationally prohibitive, we observe that $|\delta \omega_\text{iso}|$ becomes smaller with decreasing $n_\text{Th}$. This would reduce the spread in  isomer shifts between different hosts, so that our 80 MHz  scanning window width in Eq.~\eqref{Eq:Solid} can be considered as an upper bound estimate.

Finally, we point out a possible application of our developed formalism, Eq.~\eqref{Eq:VB-iso-konec},  to the {\em ab initio} determination of temperature-dependent isomer shifts in solid-state nuclear clocks. Such modeling would be important for identifying nuclear clock materials with low $T$-sensitivity (\`{a} la the 18th century temperature-compensated balance springs made from bimetallic strips).  Increasing $T$ leads to the thermal expansion, shifting equilibrium positions of the lattice nuclei. The DFT codes can be run for a thermally expanded lattice to find the $T$-dependence of electronic IPDOS$_\ell$, so that $d(\delta E_\text{iso})/dT$ can be determined from Eq.~\eqref{Eq:VB-iso-konec}. While such modeling is outside the scope of this work, we notice that in \thcaf{}  isomer shift is reduced by 0.776(3) MHz when the crystal $T$ is increased from 150 K to 293 K, as measured in ~\cite{Higgins2025-Temperature-dependence-ThCaF2}.  This variation is well below our suggested 80 MHz scanning window width and does not affect our prediction~\eqref{Eq:Solid}.

The main results of our work are the predicted values of nuclear clock 
frequencies (\ref{Eq:Solid}--\ref{Eq:ThBare}) and isomer shifts for various materials listed in Table~\ref{Tab:IPDOS-and-iso}. These are anticipated to guide  high resolution \thor{} nuclear spectroscopy in a variety of solid-state and trapped ion nuclear clock platforms, facilitating advances in nuclear timekeeping and its fundamental physics applications. 

{While the manuscript was under review, the \thor{} nuclear clock transition frequency was measured in a ThO$_2$ crystal~\cite{elwell2025-tho2}. The result of this measurement,  $2020407.5(2)_\textrm{stat}(30)_\textrm{sys}$~GHz, is consistent with our prediction, Eq.~\eqref{Eq:Solid}.}

\section{Acknowledgment}
We thank Anastassia Alexandrova, Ricky Elwell, James Terhune, and Chuankun Zhang for discussions.
This work was supported by NSF awards PHYS-2013011, PHY-2207546, and PHY-2412869 and ARO award W911NF-11-1-0369.
ERH acknowledges institutional support by the NSF QLCI Award OMA-2016245.
This work used Bridges-2 at Pittsburgh Supercomputing Center through allocation PHY230110 from the Advanced Cyberinfrastructure Coordination Ecosystem: Services \& Support (ACCESS) program, which is supported by National Science Foundation grants \#2138259, \#2138286, \#2138307, \#2137603, and \#2138296.

\section{Supplementary Material}

\subsection{Derivation of the valence band contribution to isomer shift}
{
In mean-field methods, such as the density functional theory (DFT), one solves the eigenvalue equation for
Bloch functions $\Psi_{\boldsymbol{k}}(\boldsymbol{r})$ and energy band functions $\varepsilon
_{\boldsymbol{k}}$: $H\Psi_{\boldsymbol{k}}(\boldsymbol{r})=\varepsilon
_{\boldsymbol{k}}\Psi_{\boldsymbol{k}}(\boldsymbol{r})$. Here the crystal momentum
 $\boldsymbol{k}$ varies over the Brillouin zone~\cite{callaway1991Book}.
The valence band (VB) many-body wave-function reads
$
|\text{VB}\rangle=\prod_{i \in \mathrm{VB}} c_i^\dagger \left|\tilde{0}\right\rangle 
$\,,
with $c_i^{\dagger}$ being fermionic creation operators.  We suppressed the electron spin projection $m_s$ i.e., the index $i$ stands for
sets of quantum numbers $\{\boldsymbol{k}_i, m_s^i\}$. The VB is fully occupied in an insulator. The quasi-vacuum state $\left|\tilde{0}\right\rangle$ encapsulates the remaining occupied bands.

In the second quantization, the ``dressed'' nuclear perturbation is
$
\delta \hat{\tilde{V}}=\sum_{i j} :\!c_i^\dagger c_j\!: \delta \tilde{V}_{ij}
$, 
where $\delta \tilde{V}_{ij}=\langle\Psi_{\boldsymbol{k}_i}|\delta \tilde{V}| \Psi_{\boldsymbol{k}_j}\rangle$ and $:\cdots:$ stands for the normal product of operators with respect to the quasi-vacuum~\cite{FetWal71}. Then the VB isomer shift includes contributions from all the VB electronic states 
$$
\delta E_\mathrm{iso} (\text{VB})= \langle \text{VB} | \delta \hat{\tilde{V}} | \text{VB} \rangle = 2 \sum_{\boldsymbol{k}_i}  \langle\Psi_{\boldsymbol{k}_i}|\delta \tilde{V}| \Psi_{\boldsymbol{k}_i}\rangle \,.
$$ 
Here factor of 2 accounts for the electron spin degeneracy.

Since the matrix elements  of $\delta \tilde{V}$ is accumulated near the Th nucleus ($r \lesssim a_0/Z$), we peel off the Th-centered contribution to the Bloch wavefunctions,
\begin{equation}
\Psi_{\boldsymbol{k}}(\boldsymbol{r})=\Phi_{\boldsymbol{k}}\left(
\boldsymbol{r}|\mathrm{Th}^{3+}\right)  e^{i\boldsymbol{k\cdot r}}+\ldots,
\end{equation}
by  expanding  it over the $\mathrm{Th}^{3+}$ ion atomic valence
orbitals:  $\Phi_{\boldsymbol{k}}\left(  \boldsymbol{r}%
|\mathrm{Th}^{3+}\right)  =\sum_{n\ell m}c_{n\ell m}\left(  \boldsymbol{k}\right)
\phi_{n \ell m}\left(  \boldsymbol{r}|\mathrm{Th}^{3+}\right)$, with  
$c_{n\ell m}\left(  \boldsymbol{k}\right) = \langle \phi_{n \ell m}|  \Psi_{\boldsymbol{k}} \rangle$.  To facilitate the connection with the periodic DFT numerical packages,  we limit the expansion to the lowest-energy valence states
for each partial wave, i.e., to the $7s,7p,6d,5f, \ldots$ orbitals of $\mathrm{Th}^{3+}$. This approximation is justified by the electron density near the nucleus getting progressively suppressed  with increasing $n$. Computations of similar quantities in heavy atoms~\cite{BelSafDer06,PorBelDer10} indicate a $\sim 10\%$ basis truncation error, which is comparable to the $\sim 30\%$ uncertainty in $\Delta R$. Then, 
\begin{align}
\delta E_\mathrm{iso} (\text{VB})=  
2 \sum_{\boldsymbol{k}_i}  \sum_{\ell m}   |c_{\ell m}\left(  \boldsymbol{k}_i\right)|^2 
\langle \phi_{\ell m}| \delta \tilde{V}|  \phi_{\ell m} \rangle \,, \label{Eq:VB-iso-prom}
\end{align}
where $\ell m$ refers to the lowest energy valence orbital of Th$^{3+}$ for a fixed $\ell$.

Identify $\langle \phi_{\ell m}| \delta \tilde{V}|  \phi_{\ell m} \rangle = \varepsilon_{\ell}^\mathrm{iso}(\thor^{3+})$, c.f. Eq.~(2) of the main text. Then,
\begin{align*}
\delta E_\mathrm{iso} (\text{VB})=  
2 \sum_{\boldsymbol{k}_i}  \sum_{\ell m}   |c_{\ell m}\left(  \boldsymbol{k}_i\right)|^2 
\varepsilon_{\ell}^\mathrm{iso}(\thor^{3+}) \,. 
\end{align*}
Converting the summation over discrete  set of wave-vectors to an integration over Brillouin zone (BZ)~\cite{callaway1991Book} leads to
\begin{align}
\delta E_\mathrm{iso} (\text{VB}) = 
\sum_\ell & \varepsilon_{\ell}^\mathrm{iso}(\thor^{3+}) \times \nonumber
\\ 
&\frac{2\Omega}{(2\pi)^3}  \int_\text{BZ} d^3 k \sum_{m}   |c_{\ell m}\left(  \boldsymbol{k}\right)|^2  
\,, 
\label{SI:Eq:VB-iso-prom2}
\end{align}
where $\Omega$ is the coordinate-space supercell volume (it cancels out in the final result). 

To simplify this expression, recall that 
the conventional density of states (DOS) function  is defined as~\cite{callaway1991Book}
\begin{equation}
\rho\left(  \varepsilon\right)  =\frac{2\Omega}{\left(  2\pi\right)  ^{3}}%
\int_{\text{BZ}} \,d^{3}k \, \delta\left(  \varepsilon-\varepsilon_{\boldsymbol{k}}\right) \,,
\end{equation}
where the factor of 2 comes from the electron spin degeneracy. The projected DOS (PDOS) for a partial wave $\ell$ is introduces in a similar way,
\begin{equation}
\rho_\ell\left(  \varepsilon\right)  =\frac{2\Omega}{\left(  2\pi\right)  ^{3}}%
\int_{\text{BZ}} \,d^{3}k \, \sum_m |c_{\ell m}\left(  \boldsymbol{k}\right)|^2   \delta\left(  \varepsilon-\varepsilon_{\boldsymbol{k}}\right). \label{SI:Eq:PDOS}
\end{equation}

To make use of the PDOS, insert the identity 
$1 = \int_{\text{VB}} d\varepsilon \, \delta(\varepsilon - \varepsilon_{\boldsymbol{k}})$ into Eq.~\eqref{SI:Eq:VB-iso-prom2} and swap the integrations over wave-vectors and $\varepsilon$. The result is 
\begin{align*}
\delta E_\mathrm{iso} (\text{VB}) &= 
\sum_\ell  \varepsilon_{\ell}^\mathrm{iso}(\thor^{3+}) \times \nonumber
\\ 
&\frac{2\Omega}{(2\pi)^3} \int_\text{VB} d\varepsilon \int_\text{BZ} d^3 k \sum_{m}   |c_{\ell m}\left(  \boldsymbol{k}\right)|^2   \delta(\varepsilon - \varepsilon_{\boldsymbol{k}}) \\ 
& = \sum_\ell  \varepsilon_{\ell}^\mathrm{iso}(\thor^{3+}) \int_\text{VB}  \rho_\ell\left(  \varepsilon\right)  \, d\varepsilon
\,.
\end{align*}
Then the VB contribution to isomer shift in  \thor{} solid-state clocks reads
\begin{equation}
\delta E_\mathrm{iso} (\text{VB})=
 \sum_\ell \text{IPDOS}_\ell  \times  \delta \varepsilon_\ell^\mathrm{iso}(\thor^{3+})  
 \,. \label{Eq:VB-iso-konec}
\end{equation}
Here $\text{IPDOS}_\ell  \equiv \int_\text{VB} \rho_\ell(\varepsilon) d \varepsilon$ is the integrated projected density of states (IPDOS) for a partial wave $\ell$;  the  integration extends from the VB minimum to the VB maximum.  

}

\subsection{Isomer shifts for closed-shell \thor{} ions}

The isomer shift (also known as the chemical shift) arises from the electrostatic interaction between the electron density and the nuclear charge density, which alters when the nucleus transitions between its ground and excited states. The conventional way of quantifying nuclear charge distributions is to use the root mean square (r.m.s.) radius of the nuclear charge distribution $R$. In our calculations,
we use the Fermi distribution, 
\begin{equation}
\rho_{\mathrm{nuc}}(r)=\frac{\rho_{0}}{1+\exp\left[  (r-c)/a\right]  }\,,
\label{Eq:NuclearDistros:FermiDistro}
\end{equation}
where $\rho_{0}$ is  the normalization constant, $c$ is the nuclear radius cutoff that has a meaning of the half-density ($50 \%$ fall off) radius and $a$ parameterizes  surface thickness $t=4 \ln(3) a$. We employ the typical value of $a\approx 0.52 \, \mathrm{fm}$ or, equivalently, $t\approx 2.3 \, \mathrm{fm}$. One can compute the r.m.s. radius $R$ of the Fermi distribution and relate it to $R$, see e.g.~\cite{WRJBook}. In particular, $\delta \langle r^2 \rangle^{229m,229} = 0.0097(26) \, \mathrm{fm}^2$ from Ref.~\cite{Yamaguchi2024}, used in the main text, translates into the difference $\Delta c = 0.00116(31)\,\mathrm{fm}$ between the half-density radii of the nuclear isomer and ground states.

We run the {\em ab initio} relativistic Dirac-Hartree-Fock code with 
several values of $c$, obtain the total electronic energy $E(c)$ and determine the numerical derivative  ${d  E(c)}/{d c}$. Then the isomer shift is computed as  
\begin{equation}
\delta E_\mathrm{iso} = \frac{d  E(c)}{d c} \, \Delta c\,.
\label{isomarEq}
\end{equation}
Our numerical results are compiled in Table~\ref{SI:Tab:isomer-shift-Th-ions}.

\begin{table}[h!] 
\caption{The isomer shift derivatives $dE(c)/dc$ , and the isomer-shifts $\delta E_\mathrm{iso} $ for the closed-shell Th ions with  their noble gas-like electronic configurations listed in square brackets. These isomer shifts are with respect to the bare nucleus, \thor$^{90+}$.} 
\label{diff-thions} 
\begin{ruledtabular} 
\begin{tabular}{ccc} 
Ion & $dE/dc$, a.u./fm & $\delta E_\mathrm{iso}$, eV \\ 
\hline 
$\thor^{4+}$ [Rn]  & 2.644  & 0.0833 \\ 
$\thor^{36+}$ [Xe] & 2.650  & 0.0838 \\ 
$\thor^{54+}$ [Kr] & 2.647  & 0.0836 \\ 
$\thor^{72+}$ [Ar] & 2.631  & 0.0829 \\ 
$\thor^{80+}$ [Ne] & 2.522  & 0.0797 \\ 
$\thor^{88+}$ [He] & 2.142  & 0.0675  
\label{SI:Tab:isomer-shift-Th-ions} 
\end{tabular} 
\end{ruledtabular} 
\end{table}

The isomer shifts in Table~\ref{SI:Tab:isomer-shift-Th-ions}  are dominated by the $1s^2$ shell  having the largest overlap with the nucleus. 
To verify the numerical results,  we computed the isomer shift  analytically for He-like 
$^{229}$Th$^{88+}$ ion using the uniform nuclear charge distribution. In our estimates, we used the effective charge $Z_\mathrm{eff} = Z-{5}/{16}$ to account for electronic correlations in the $1s^2$ shell~\cite{WRJBook} and the H-like Dirac wavefunctions with $Z_\mathrm{eff}$ for a point-like nucleus. The first-order correction in the electrostatic nuclear potential yields 
$0.0642\, \text{eV}$. This value is consistent with the numerical finite-size nucleus result of $0.0675\, \text{eV}$  in Table~\ref{SI:Tab:isomer-shift-Th-ions}.

Each of the listed isomer shifts in Table~\ref{SI:Tab:isomer-shift-Th-ions} has an error of at least 27\% due to the uncertainty in $\delta \langle r^2 \rangle^{229m,229} = 0.0097(26) \, \mathrm{fm}^2$. In particular, 
\begin{equation}
    \delta E_\mathrm{iso}(\thor^{4+}) =  0.083(22) \, \text{eV} \,.
\end{equation}

\subsection{Isomer shifts for the $^{229}$Th$^{3+}$ ion valence states}
\begin{table}[h!]
\caption{Isomer shifts for the valence states of $\thor^{3+}$. The notation $x[y]$ stands for $x \times 10^y$.}
\label{tab:example}
\begin{ruledtabular}
 \centering
\begin{tabular}{ccc}
State &  $\delta ^\mathrm{iso}_{n\ell j}$, MHz, Ref.~\cite{DzubaFlam2023-ThIsomerShift}  & $\delta \varepsilon_{n\ell j}^\mathrm{(1),iso}$, MHz  \\
\hline
$7s_{1/2}$  &8.93[2]&5.93[2]\\
$7p_{1/2}$ &2.62[1]&4.77[1]\\
$7p_{3/2}$ &-5.14[1]&9.62[-04]\\
$6d_{3/2}$ &-2.26[2]&3.70[-05]\\
$6d_{5/2}$  &-1.99[2]&7.60[-10]\\
$5f_{5/2}$ &-5.34[2]&3.99[-12]\\
$5f_{7/2}$ &-5.14[2]&6.48[-17]\\
\end{tabular}
\end{ruledtabular}
\label{isomar_final}
\end{table}

Table~\ref{isomar_final} compiles isomer shifts for the $^{229}$Th$^{3+}$ lowest-energy valence states of a given $\ell_j$ symmetry. The $\delta ^\mathrm{iso}_{n\ell j}$ values are from Ref.~\cite{DzubaFlam2023-ThIsomerShift}, which used a  combination of the Dirac-Hartree-Fock (DHF) method and the random-phase approximation (RPA). Our DHF+RPA-like calculations, carried out by running DHF code for different values of the nuclear radius $R$, are in a good agreement with Ref.~\cite{DzubaFlam2023-ThIsomerShift} values.  We computed the $\delta \varepsilon_{n\ell j}^\mathrm{(1),iso}$ values by evaluating first-order correction in $\delta V$ to the energy of the Th$^{3+}$  valence electron with the DHF method. The comparison indicates that for a fixed valence state, the isomer shifts can differ substantially due to the relaxation effect elucidated in the main text. The agreement becomes especially pronounced (many orders of magnitude) for the states with a small overlap with the nucleus, i.e. $j>1/2$ states.  
Notably, for the $5f_{7/2}$ orbital, the computed isomer shifts  differ by 19 orders of magnitude  and exhibit opposite signs.

\subsection{DFT methods}
DFT calculations on were performed with VASP~\cite{RN12}, using the PAW~\cite{RN14} method with a plane-wave cutoff of 500 eV and a spin-restricted formalism.
VASP version 6.3 was used for Th:\ce{LiSrAlF6} and Th:\ce{CaF2}, and version 6.4.2 for \ce{ThF4}, Th:\ce{BaMgF4}, and Th:\ce{BaZnF4}.
The PBE\cite{RN13} functional was used for all structural optimizations, and the modified Becke-Johnson (MBJ)\cite{RN489,RN490} functional was used for electronic properties.
Further particulars of the calculations can be found in papers in which each material was studied in depth.\cite{Elwell2024,zhang2024thf,Morgan2025-Th229NonLinear}
Numerical integration of the PDOS to compute the IPDOS was done with the \textit{numpy} implementation of trapezium integration.

We tested the sensitivity of the IPDOS to three computational parameters - the $k$-point grid, the plane-wave cut-off, and the exchange-correlation functional.
\ce{ThF4} was used for these test calculations because of its smaller unit cell.
We also tested the effect of recalculating the thorium IPDOSs in \ce{ThF4} in a $1 \times 1 \times 2$ supercell to see if increasing the supercell size without changing the structure in any way affects the IPDOS.
When other variables were being tested, the $k$-point grid spacing was 0.02 \AA{}$^{-1}$, the plane-wave cut-off was 500 eV, and the functional was PBE.

\begin{figure}
    \centering
    \includegraphics[width=\linewidth]{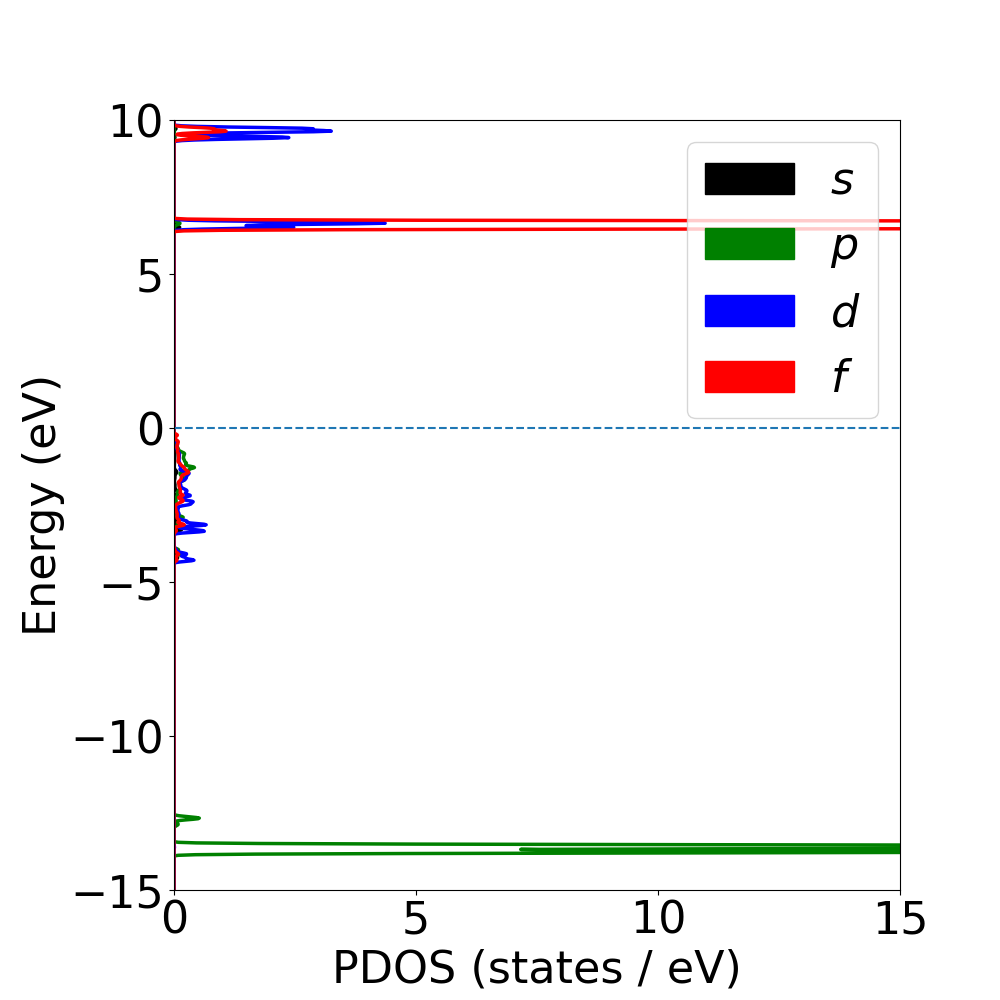}
    \caption{Thorium PDOS for Th:\ce{LiSrAlF6} showing the Th $6p$ states.}
    \label{fig:Th:Lisaf semicore PDOS}
\end{figure}

To compute the IPDOS for a specific compound it is necessary to define the energy range for the integration.
The densities of states for these thorium fluorides have a common structure to the states below the Fermi level; the ``valence band'' sits immediately below the Fermi level and is composed of fluorine $2p$ orbitals with a small admixture of thorium orbitals, below which there is a large gap with no states, and then the Th $6p$ semi-core states appear.
This is illustrated for Th:\ce{LiSrAlF6} in Figure \ref{fig:Th:Lisaf semicore PDOS}.
The valence band runs from 0 to -5 eV, and the Th $6p$ peaks appear at approximately -14 eV.
We integrate the IPDOS over the valence band only, in this case from -10 to 0 eV, so the $6p$ states are not included in the IPDOS.
This is because the Th $6p$ electrons are described by the \ce{Th^4+} core in the DHF calculations.
It is necessary to locate the $6p$ orbitals for each material in this way before computing the IPDOS.
Integration of the total DOS over the Th $6p$ peak in Th:\ce{LiSrAlF6} gives exactly 6 electrons, indicating that the $6p$ electrons are correctly treated as valence electrons by the DFT calculations and as core electrons by the DHF calculations.
The DFT IPDOSs for $\ell \neq 1$ sub-shells are $s= 0.001$, $d= 0.002$, and $f= 0.001$, indicating that the Th $6p$ peak contains negligible contributions from other orbitals.

\

The results are shown in the tables below.
Note that ``Th 1'' and ``Th 2'' are the two symmetry-inequivalent thorium atoms in the unit cell of \ce{ThF4}.
In our notation the cell contains four atoms of type 1 and two of type 2.
The IPDOSs are converged with respect to $k$-point sampling at all tested grid spacings, and with respect to plane-wave cut-off at 500 eV (the value used in all our calculations) if not lower.
The IPDOSs are, perhaps unsurprisingly, sensitive to the choice of DFT functional.
PBE and PBEsol, both GGA functionals, give the highest Th IPDOSs.
Adding a Hubbard U correction of 4 eV to the $f$ orbitals reduces the $f$ IPDOS but the other orbitals are unaffected.
MBJ, a meta-GGA functional specialized for electronic properties and band gaps, gave the smallest IPDOSs.
MBJ is the functional used for the calculations reported in the main text.
R2SCAN, another meta-GGA functional, gave IPDOSs intermediate between MBJ and PBE.
HSE06, a hybrid functional, gave results that closely match MBJ, especially for $f$ and $d$ orbitals. {Notice that the {\em differential} isomer shift between the two sites in \ce{ThF4}, given as $\Delta\delta\omega_\mathrm{iso}^{12}$ in Table~\ref{IPDOS functional test}, is largely independent of the choice of the functional. We obtain a range from 3 to 5 MHz.}

Unfortunately, it is computationally unfeasible to do hybrid functional calculations on the large unit cells needed to study thorium defects in doped hosts.
Doubling the unit cell of \ce{ThF4} from $1 \times 1 \times 1$ to $1 \times 1 \times 2$ did not change the Th IPDOSs.
This supercell transformation does not change the chemical structure being described in any way, so all predicted properties should remain the same if the calculations are behaving correctly.
This result implies that the sensitivity to supercell size seen in Th:\ce{LiSrAlF6} is due to Th-Th interactions between periodic images in the doped supercells.

\begin{table}[]
\begin{tabular}{cccccc}
\hline
\hline
Cut-off, eV & $s$   & $p$   & $d$   & $f$ & $\delta \omega_\mathrm{iso}$, MHz   \\
\hline
Th 1         &       &       &       &      & \\
300          & 0.108 & 0.481 & 0.660 & 0.456  & -292\\
400          & 0.108 & 0.482 & 0.657 & 0.453 & -290\\
500          & 0.108 & 0.481 & 0.660 & 0.456 & -292\\
600          & 0.108 & 0.481 & 0.659 & 0.456 & -292\\
700          & 0.108 & 0.481 & 0.659 & 0.455 & -292 \\
800          & 0.108 & 0.481 & 0.659 & 0.455 & -292\\
\hline
Th 2         &       &       &       &       \\
300          & 0.107 & 0.469 & 0.654 & 0.451 & -289 \\
400          & 0.107 & 0.470 & 0.651 & 0.448 & -287 \\
500          & 0.107 & 0.469 & 0.654 & 0.451 & -289 \\
600          & 0.107 & 0.469 & 0.653 & 0.451 & -289\\
700          & 0.107 & 0.469 & 0.653 & 0.450 & -288\\
800          & 0.107 & 0.469 & 0.653 & 0.450 & -288 \\
\hline
\hline
\end{tabular}
\caption{Th IPDOSs and isomer shifts  for \ce{ThF4} calculated using a range of plane-wave cut-off energies}
\label{IPDOS cut-off test}
\end{table}

\begin{table}[]
\begin{tabular}{lccccc}
\hline
\hline
$k$-point spacing, \AA{}$^{-1}$ & $s$   & $p$   & $d$   & $f$  & $\delta \omega_\mathrm{iso}$, MHz   \\
\hline
Th 1                      &       &       &       &       &\\
0.03                      & 0.108 & 0.481 & 0.660 & 0.456 & -292\\
0.025                     & 0.108 & 0.481 & 0.660 & 0.456 & -292\\
0.02                      & 0.108 & 0.481 & 0.660 & 0.456 & -292 \\
0.015                     & 0.108 & 0.481 & 0.660 & 0.456 & -292 \\
0.0125                    & 0.108 & 0.481 & 0.660 & 0.456  & -292\\
\hline
Th 2                      &       &       &       & &\\
0.03                      & 0.107 & 0.469 & 0.654 & 0.451 & -289 \\
0.025                     & 0.107 & 0.469 & 0.654 & 0.451 & -289 \\
0.02                      & 0.107 & 0.469 & 0.654 & 0.451 & -289 \\
0.015                     & 0.107 & 0.469 & 0.654 & 0.451 & -289\\
0.0125                    & 0.107 & 0.469 & 0.654 & 0.451 & -289\\
\hline
\hline
\end{tabular}
\caption{Th IPDOSs and isomer shifts for \ce{ThF4} calculated using a range of $k$-point grids}
\label{IPDOS kpoints test}
\end{table}

\begin{table}[]
\begin{tabular}{ccccccc}
\hline
\hline
Functional & $s$   & $p$   & $d$   & $f$ & $\delta \omega_\mathrm{iso}$, MHz & $\Delta\delta\omega_\mathrm{iso}^{12}$\\
\hline
Th 1       &       &       &       &     &  & \\
PBE        & 0.108 & 0.481 & 0.660 & 0.456 & -292 & \\
PBE+U      & 0.112 & 0.472 & 0.687 & 0.338 & -233 &  \\
PBEsol     & 0.111 & 0.489 & 0.672 & 0.459 & -294 &  \\
PBEsol+U   & 0.114 & 0.479 & 0.699 & 0.341 & -235 &  \\
MBJ        & 0.074 & 0.424 & 0.599 & 0.382 & -270 &  \\
R2SCAN     & 0.095 & 0.477 & 0.61  & 0.42  & -275 &  \\
HSE06      & 0.102 & 0.523 & 0.616 & 0.378 & -249 &  \\
\hline
Th 2      &       &       &       & &\\
PBE        & 0.107 & 0.469 & 0.654 & 0.451 & -289 & 3 \\
PBE+U      & 0.111 & 0.46  & 0.68  & 0.333 & -229 & 4 \\
PBEsol     & 0.11  & 0.477 & 0.666 & 0.453 & -290 & 4 \\
PBEsol+U   & 0.114 & 0.468 & 0.693 & 0.334 & -230 & 5 \\
MBJ        & 0.073 & 0.413 & 0.593 & 0.376 & -266 & 4 \\
R2SCAN     & 0.094 & 0.465 & 0.604 & 0.415 & -271 & 4 \\
HSE06      & 0.101 & 0.511 & 0.61  & 0.373 & -246 & 3 \\
\hline
\hline
\end{tabular}
\caption{Th IPDOSs for \ce{ThF4} calculated using a range of DFT functionals}
\label{IPDOS functional test}
\end{table}

\subsection{Supercell sizes used for periodic DFT calculations}

Our periodic DFT model requires that the thorium defect be described in a supercell of the host crystal.
The sizes of all supercells used in this study (multiplicity and lattice vectors in \AA{}) are given in the  table \ref{Supercell size table}.
\ce{ThF4} is included, but note that this is not a doped compound so the unit cell (\textit{i.e.} a $1 \times 1 \times 1$ supercell) was used for the IPDOS calculation.

\begin{table}[h]
\begin{tabular}{ccccc}
\hline
\hline
       & supercell & $a$ (\AA{})  & $b$ (\AA{})  & $c$ (\AA{})  \\
       \hline
\ce{LiSrAlF6} & 3x3x2     & 15.461 & 15.461 & 20.714 \\
\ce{LiSrAlF6} & 4x4x2     & 20.615 & 20.615 & 20.714 \\
\ce{CaF2}   & 3x3x3     & 16.492 & 16.492 & 16.492 \\
\ce{ThF4}   & 1x1x1     & 8.605  & 8.605  & 8.605  \\
\ce{BaMgF4} & 4x1x3     & 16.560 & 14.835 & 17.680 \\
\ce{BaZnF4} & 4x1x3     & 16.895 & 14.874 & 17.799 \\
\hline
\hline
\end{tabular}
\caption{Supercell sizes for DFT calculations of thorium-containing crystals.}
\label{Supercell size table}
\end{table}


\end{document}